\documentstyle[12pt]{article}

\def\fun#1#2{\lower3.6pt\vbox{\baselineskip0pt\lineskip.9pt
        \ialign{$\mathsurround=0pt#1\hfill##\hfil$\crcr#2\crcr\sim\crcr}}}

\renewcommand\({\left(}
\renewcommand\){\right)}

\newcommand\eq[1]{Eq.~(\ref{#1})}
\newcommand\eqs[2]{Eqs.~(\ref{#1}) and (\ref{#2})}

\newcommand\ee{\end{equation}}
\newcommand\be{\begin{equation}}
\newcommand\eea{\end{eqnarray}}
\newcommand\bea{\begin{eqnarray}}

\newcommand\GeV{\,\mbox{GeV}}
\newcommand\MeV{\,\mbox{MeV}}

\newcommand\mpl{M_{\rm P}}

\newcommand\mpsis{|m_\psi^2|}
\newcommand\etapsi{\eta_\psi}
\newcommand\luv{\Lambda_{\rm UV}}

\newcommand\lsim{\mathrel{\rlap{\lower4pt\hbox{\hskip1pt$\sim$}}
    \raise1pt\hbox{$<$}}}
\newcommand\gsim{\mathrel{\rlap{\lower4pt\hbox{\hskip1pt$\sim$}}
    \raise1pt\hbox{$>$}}}

\def\dslash{\not{\hbox{\kern-2pt $\partial$}}}
\def\Dslash{\not{\hbox{\kern-4pt $D$}}}
\def\Oslash{\not{\hbox{\kern-4pt $O$}}}
\def\Qslash{\not{\hbox{\kern-4pt $Q$}}}
\def\pslash{\not{\hbox{\kern-2.3pt $p$}}}
\def\kslash{\not{\hbox{\kern-2.3pt $k$}}}
\def\qslash{\not{\hbox{\kern-2.3pt $q$}}}

 \newtoks\slashfraction
 \slashfraction={.13}
 \def\slash#1{\setbox0\hbox{$ #1 $}
 \setbox0\hbox to \the\slashfraction\wd0{\hss \box0}/\box0 }
 
\def\ee{\end{equation}}
\def\be{\begin{equation}}

\newcommand\sub[1]{_{\rm #1}}

\begin{document}

\begin{flushright}
LANCS-TH/9823
\\hep-ph/9904371\\
(April 1999)
\end{flushright}
\begin{center}
{\Large \bf The parameter space for tree-level hybrid inflation}

\vspace{.3in}
{\large\bf  David H.~Lyth}

\vspace{.4 cm}
{\em Department of Physics,\\
Lancaster University,\\
Lancaster LA1 4YB.~~~U.~K.}
\vspace{.4cm}
\end{center}

\begin{abstract}
In a large region of parameter space, the tree-level hybrid inflation 
model is likely to be be invalidated by loop corrections. 
In particular, this is likely to be the case if both quartic couplings
are of order unity, as is often supposed. It is likely also to be the 
case if there is an ultra-violet cutoff far below the Planck scale,
($\luv\lsim 10^9\GeV(V_0^{1/4}/1\MeV)^{2/5}$, where $V_0$ is the height
of the potential) 
unless one allows field values bigger than the cutoff.
\end{abstract}

1. The original tree-level 
hybrid inflation model \cite{andhyb} involves the mass of the inflaton
field $\phi$,
the coupling $\lambda'\phi^2\psi^2$ of the inflaton to another
field $\psi$, and the self-coupling $\lambda\psi^4$.
It was immediately noted that the model can give
a density perturbation of the required magnitude with
the natural choice $\lambda\sim\lambda'\sim 1$.
This feature, desirable in itself, yields the
bonus \cite{cllsw} that all relevant field values are
far below the Planck scale,
making it feasible to justify the neglect of non-renormalizable terms
in the potential
\cite{ewansg,treview}.

Supersymmetric realizations of the tree-level model
were soon proposed \cite{cllsw,ewansg}, 
which seemed to confirm its theoretical viability.
Unfortunately, with
supersymmetry in place it became sensible to evaluate the 
1-loop correction to the potential.
In both models, the loop corrections was found
\cite{giaf,giad}
to be large, necessarily dominating
the tree-level term if both couplings are of order 1.\footnote
{One of the models had the 
$F$ term dominating \cite{cllsw,ewansg} while the other
had the $D$ term dominating \cite{ewansg}.
The models actually made the 
potential completely flat during inflation, but supergravity corrections 
generate \cite{cllsw,ewansg}
a non-zero (in fact generically rather too large) curvature
for the potential. The loop correction was actually first evaluated
\cite{giaf,giad} 
for slight variants of the original models.}

In this note I argue that the loop correction in a generic model will
be at least as big as the one found in these particular cases, unless 
there are unforseen cancellations between unrelated parameters.
On this basis, I describe the maximal region of parameter space in which the 
tree-level model can be valid. As already indicated, it
does not include the regime of unsuppressed couplings,
which unfortunately is still in common use.\footnote
{The two most recent examples are \cite{andrei} which invokes this 
regime to construct a model of inflation with TeV-scale quantum gravity,
and \cite{juan} which uses it to construct a model of baryogenesis.}
I shall quote without comment some 
well-known results, referring for details to a recent review 
of inflation \cite{treview}.

2. The original tree-level hybrid inflation model \cite{andhyb}
is defined by
\be
V(\phi,\psi)= V_0 +\frac12m^2\phi^2 +\frac12\lambda'\psi^2\phi^2
-\frac12\mpsis\psi^2 + \frac14\lambda\psi^4 \,.
\label{fullpot}
\ee
Inflation takes place in the 
regime $\phi^2>\phi\sub c^2\equiv \mpsis/\lambda'$, where
$\psi$ is driven to zero and the inflaton potential is
\be 
V=V_0+\frac12m^2\phi^2 \,. \label{vofphi}
\ee
The constant term $V_0$ is assumed to dominate during inflation.

The last term of \eq{fullpot} serves only to determine the vacuum 
expectation value (vev) of $\psi$, achieved when $\phi$ falls below
$\phi\sub c$. Using
that fact that $V_0$ vanishes in the vacuum, one learns that
the vev is
\be
\langle\psi\rangle\equiv
M= 2 V_0^{1/2}/|m_\psi|=\lambda^{-1/2} |m_\psi| ,.
\ee
If the last term of \eq{fullpot} is replaced by
a higher (but not huge) power of $\psi$ one still has
roughly
\be
M\sim 2 V_0^{1/2}/|m_\psi| \,,
\ee
and for the most part I will invoke only this result,
making no explicit use of the parameter $\lambda$.
The results will therefor not be sensitive to the precise power
appearing in the the last term of \eq{fullpot}.

During inflation, it
is useful to define parameters
\bea
\eta &\equiv& \frac{m^2 \mpl^2}{V_0} \\
\etapsi &\equiv & \frac{\mpsis\mpl^2}{V_0} \sim \frac{4\mpl^2}{M^2}
\label{etapsiapp}
\,.
\eea
As usual, $\mpl\equiv (8\pi G)^{-1/2}
=2.4\times 10^{18}\GeV$ is the Planck scale.
Slow-roll inflation requires $\eta\ll 1$, and a prompt end to inflation at
$\phi\sub c$ requires $\etapsi\gg1$. The latter condition 
can be interpreted loosely \cite{rsg}, but
the former has to be tight because the spectral index $n$ is predicted 
as $n-1=2\eta$ whereas observation requires $|n-1|\ll 1$.
A preliminary estimate from recent observations is \cite{bond}
$|n-1|<0.05$, so it seems reasonable to assume 
\bea
\eta &\lsim& 0.025 
\label{etabound}\\
\etapsi &\gsim& 1
\eea

The scales probed by COBE leave the horizon $N<60$ $e$-folds
before the end of slow-roll inflation, when the field is
\be
\phi^2 =e^{2\eta N} \phi\sub c^2 \simeq \phi\sub c^2 
\equiv \mpsis/\lambda' \,.
\label{phi}
\ee
(The approximate equality corresponds to $e^{2\eta N}\sim 1$
which is good enough for order-of-magnitude estimates
since $\eta<0.025$ requires $e^{2\eta N}\lsim 5$.)
The COBE normalization of the spectrum of the density perturbation
gives the constraint \cite{andhyb,cllsw,treview}
\bea
\lambda' &=& 2.8\times 10^{-7} e^{2\eta N} \eta^2 \eta_\psi \\
&\simeq& 3 \times 10^{-7} \eta^2 \eta_\psi \,.
\eea
After imposing this constraint, the requirement that $V_0$ dominates
 the inflaton potential becomes
\be
\frac{V_0}{\mpl^4} \ll 3\times 10^{-7}\eta \,.
\label{vzerodom}
\ee

3. The 1-loop correction to the inflationary potential 
coming from $\psi$ is
\be
\Delta V = \frac1{32\pi^2} m_\psi^4(\phi)
\ln\(\frac{m_\psi(\phi)} Q \) \,
\label{delvfull}
\ee
where 
\be
m_\psi^2(\phi) \equiv \(\lambda'\phi^2-\mpsis \)
=\lambda'(\phi^2-\phi\sub c^2)
 \,.
\ee
(A constant of order 1 has been dropped in the argument of the log,
which depends on the renormalization scheme and does not affect the 
conclusions.) In this expression, 
$Q$ is the renormalization scale at which the parameters of the 
tree-level potential should be evaluated. Its choice is arbitrary,
and if all loop corrections were included
the total potential would be independent
of $Q$. In any application of quantum field theory, one 
should choose $Q$ so that the total 1-loop correction is
small, hopefully justifying the neglect of multi-loop correction.
In the present context this is achieved by choosing
\be
Q\sim \sqrt{\lambda'} \phi\sub c \,.
\ee

In order to justify the omission of a quartic term $\propto \phi^4$
in the tree-level potential,
and also to protect the parameters against 
radiative corrections, one assumes that the underlying theory
is supersymmetric. The above loop correction
will then be accompanied
by the loop corrections coming from the spin-zero and spin-half
superpartners of
$\psi$.
If supersymmetry were unbroken, the total loop correction would
vanish, but supersymmetry is necessarily broken during inflation.
(The scale of susy breaking, defined essentially as the magnitude
of the auxiliary field(s) breaking susy, cannot be less than
$V^{1/2}$.) The breaking can be either spontaneous or soft,
but in either case the superpartners couple to $\phi$ with the
{\em same} coupling strength $\lambda'$.
The total loop correction from $\psi$ and its superpartners
is\footnote
{The mass matrix is taken to be diagonal for simplicity, since 
the non-diagonal case does not seem to introduce any essentially new
feature.}
\be
\Delta V = \frac1{32\pi^2} \(
m_\psi^4(\phi) \ln\(\frac{m_\psi(\phi)} Q \) 
+
m_{\tilde \psi}^4(\phi) \ln\(\frac{m_{\tilde\psi}(\phi)} Q \) 
-2
m\sub f^4(\phi) \ln\(\frac{m\sub f(\phi)} Q \) 
\)
\, \,,
\label{total}
\ee
where 
\bea
m_{\tilde\psi}^2(\phi) &=& \lambda'\phi^2 + m_{\tilde\psi}^2 \\
m\sub f(\phi) &=& \sqrt{\lambda'}\phi + m\sub f \,.
\eea
Here $\tilde\psi$ is the field of the scalar partner, while f denotes
the spin-half partner. The mass $m\sub f\equiv m\sub f(0)$ vanishes
if the full potential respects the symmetry $\phi\to-\phi$ of the 
original, which we assume.

The mass-squared $m\sub{\tilde\psi}^2$ may be either positive or 
negative. In the former case, $\tilde\psi$ vanishes both during 
inflation and in the true vacuum, automatically justifying its omission in the 
tree-level potential. In the latter case, the omission 
of $\tilde\psi$ is strictly justified only if $|m_{\tilde\psi}|
=|m_\psi|$, the only effect of including it then being
the trivial replacement $\psi^2\to\psi^2+\tilde\psi^2$.
However, its omission is justified in practice also if 
$|m_{\tilde\psi}|$ is {\em roughly} of order
$|m_\psi|$. Indeed, if
$|m_{\tilde\psi}|<|m_\psi|$, the field $\tilde\psi$ is held at the 
origin until the field $\psi$ is destabilized, making the former
ineffective except for a modest
increase in the value of $V_0$. The
opposite case may be discounted because it is equivalent
to an interchange of the labels
$\psi$ and $\psi'$. In the following we therefore assume
\be
|m_{\tilde\psi}|<|m_\psi| \label{mineq}
\,.
\ee

I am going to argue that with these assumptions,
the derivatives of \eq{total} will be at least of the following order
\bea
\Delta V' \sim\frac{|m_\psi|^4}{\phi\sub c} &= &\sqrt{\lambda'} |m_\psi|^3
\label{delvp}
\\
\Delta V'' \sim \frac{|m_\psi|^4}{\phi\sub c^2} &=& \lambda' |m_\psi|^2 \,.
\label{delvpp}
\eea

Consider first the case that $\eta$ is not too small, so that
$e^{2\eta N}
$ is significantly bigger than 1. In this case,
$\lambda'\phi^2 $ is significantly bigger than $\mpsis$
(equivalently, 
$\phi$ is significantly bigger than $\phi\sub c$)
while COBE scales leave 
the horizon, and remains so until the last few $e$-folds of inflation.
As a result, the $\psi$ contribution \eq{delvfull}
is given at least roughly by
\be
\Delta V \sim \frac1 {32\pi^2} 
\(\lambda'^2\phi^4 - 2\lambda'\mpsis \phi^2 + |m_\psi|^4 \)
\ln(\sqrt{\lambda'}\phi /Q) \,.
\label{delv}
\ee
The $\phi^4$ term in front of the log disappears when the 
contributions of the superpartners are included
(\eq{total} with the arguments of the logs identical).
In models with spontaneously broken global supersymmetry 
the $\phi^2$ term will also typically be cancelled (in other words
one will typically have $m_{\tilde\psi}^2=\mpsis$).
The coefficient of the 
constant term will however {\em not} be cancelled, and keeping 
only it
gives the estimates \eqs{delvp}{delvpp}.\footnote
{In models with softly broken supersymmetry the $\phi^2$ term
will also survive and in general be dominant,
but the contribution of the constant term is still there.}

If $\eta$ is many orders of magnitude below 1, so
that $\phi$ is very close to $\phi\sub c$, \eq{delv} will not be 
a good approximation for the $\psi$ contribution.
As a result the spin-half contribution to \eq{total} is now
uncancelled (to be precise, half of it remains uncancelled while the 
other half might be cancelled by the $\tilde\psi$ contribution).
Evaluating the derivatives of this contribution again leads to estimates
of order \eqs{delvp}{delvpp}.

This justifies \eqs{delvp}{delvpp} as rough estimates
of the 1-loop contribution of $\psi$ and its superpartners,
to the derivatives of the potential.
As there is no reason why this contribution should be cancelled by
additional contributions
(coming from additional Yukawa couplings or from
gauge fields \cite{ewanloop})  \eqs{delvp}{delvpp} are also
rough lower bounds on the total 1-loop contribution.

For the tree-level model to be valid, the 
first two derivatives of the loop correction ought to be negligible,
$\Delta V'\ll m^2\phi\sub c$ and
$\Delta V''\ll m^2$. The second constraint is the strongest, and it 
may be written
\be
\frac1{32\pi^2} \lambda' \ll \frac\eta {\etapsi} \,.
\label{b1}
\ee

Using the COBE normalization we can write this constraint
in various ways.
Eliminating $\etapsi$ it becomes
\be
\lambda' \ll (\eta/22)^{3/2} \lsim 4\times 10^{-5} \,.\label{first}
\ee
We see that $\lambda'$ has to be quite small.
Eliminating instead $\eta$ we find a result that can usefully be written in 
terms of the vev $M$ of $\psi$, 
\be
\lambda' \ll (42M/\mpl)^6 \,.\label{second}
\ee
This is a stronger bound if $M/\mpl\lsim 0.005(\eta/.025)^{1/4}$.
The third simple possibility is to eliminate $\lambda'$, leading to
a result that can be written in similar form,
\be
\eta\ll (90M/\mpl)^4 \,.\label{third} 
\ee
This is more restrictive than \eq{etabound} if 
$M/\mpl\lsim 5\times 10^{-3}$.
Using \eq{vzerodom} we learn that $V_0^{1/4}\ll 1.6 M$.

We can also use \eq{second} to 
obtain a lower bound on the value \eq{phi} of $\phi\sub c$, which can be 
written
\be
\( \frac M{\mpl} \)^4 \( \frac {\phi\sub c}{\mpl} \) \gsim \( \frac{V_0^{1/4}}
{156\mpl} \)^2 \gsim \( \frac{10^{9}\GeV}{\mpl} \)^5 \,.
\label{fourth}
\ee
To obtain the last inequality we set $V_0^{1/4}\sim 1\MeV$,
the smallest possible value since reheating must occur before
nucleosynthesis.

4. The tree-level potential \eq{fullpot} ignores 
nonrenormalizable terms. They are  of the form $\lambda_{mn}\luv^{4-m-n}\phi^m
\psi^n$, with $m+n\geq 5$, where $\luv$ is the ultra-violet 
cutoff for the effective field theory relevant during inflation.
These terms summarize the physics which is ignored by the effective 
field theory.

Since quantum gravity certainly become
significant on Planckian scales one must have $\luv\lsim \mpl$,
but $\luv$ will be smaller if the effective field theory breaks
before Planckian scales are reached. This could happen in 
three ways. First,
a different field theory may take over,
containing 
fields that have been integrated out in the effective theory. Second,
field theory may give way to string theory. Third, the scale of quantum 
gravity could be lower than $\mpl$ because there are extra dimensions 
with a large compactification radius.

The coefficients $\lambda_{mn}$ are generically of 
order $1$, so their neglect is normally justified only if all field
values are $\lsim \luv$.\footnote
{In the context of global supersymmetry with minimal kinetic terms,
the holomorphy of the superpotential allows one to eliminate some of 
the non-renormalizable terms by imposing suitable internal symmetries
but they are still dangerous in the full supergravity theory
\cite{treview}.}
\eq{fourth} implies that this
is possible only if 
\be
\luv\gsim 10^{9}\GeV\( V_0^{1/4}/1\MeV\)^{2/5} \label{luvbound}
\,.
\ee

5. We noted already that our results remain valid if
the last term of \eq{fullpot} is replaced 
by a non-renormalizable term.
Let us ask what happens if 
the term 
$\frac12\lambda'
\phi^2\psi^2$ 
 is replaced by a non-renormalizable
term, specifically the term
$\phi^4\psi^2/\luv^2$  proposed by Randall et al.~\cite{rsg}.
This gives $\phi\sub c^2=\luv |m_\psi|$
instead $\phi\sub c^2=\mpsis/\lambda'$. 
\eqs{third}{fourth}
are unaffected, while \eqs{first}{second} become
\bea
\frac{V_0^{1/2}}{2M\luv} &\lsim & \(\frac{42M}{\mpl} \)^6 \\
\frac{V_0^{1/2}}{2M\luv} &\lsim & \(\frac\eta{22} \)^{3/2} 
\lsim 4\times 10^{-5}\,.
\eea
Invoking again\footnote
{In advocating this condition we differ from Randall et al.~\cite{rsg},
who 
advocate $M\sim\mpl$ even in the presence of the ultra-violet cutoff.}
the condition$M\lsim\luv$, these inequalities
become lower bounds on $\luv$.
Invoking again $V^{1/4}\gsim
1\MeV$, the first inequality  becomes $\luv\gsim 10^{12}\GeV$,
significantly stronger than \eq{luvbound}.

6. In summary, the parameter space for tree-level hybrid inflation
is strongly constrained. It does not include the regime where both
dimensionless couplings are unsuppressed. Nor does it include the regime
where all relevant field values are far below the Planck scale.
The second result is perhaps the most interesting, because it 
means that tree-level hybrid inflation is unviable if there is an
ultra-violet cutoff far below the Planck scale, unless one is willing
to allow field values much bigger than the cutoff.

\section*{Acknowledgements}
I am indebted to Toni Riotto for many useful discussions.

\newcommand\pl[3]{Phys. Lett. {\bf #1}, #2 (19#3)}
\newcommand\np[3]{Nucl. Phys. {\bf #1}, #2 (19#3)}
\newcommand\pr[3]{Phys. Rep. {\bf #1}, #2 (19#3)}
\newcommand\prl[3]{Phys. Rev. Lett. {\bf #1}, #2 (19#3)}
\newcommand\prd[3]{Phys. Rev. D{\bf #1}, #2 (19#3)}
\newcommand\ptp[3]{Prog. Theor. Phys. {\bf #1}, #2 (19#3)}

\end{document}